# Three potential sources of cell injuries of lymphoid cells associated with developments of lymphoid leukemia and lymphoma

Jicun Wang-Michelitsch[1]*, Thomas M Michelitsch[2]

[1] Independent researcher

[2] Sorbonne Université, Institut Jean le Rond d'Alembert, CNRS UMR 7190 Paris, France

**Abstract**

Lymphoid leukemia (LL) and lymphoma are two groups of blood cancers developed from lymphoid cells (LCs). To understand the cause and the mechanism of cell transformation of a LC, we studied the potential sources of cell injuries of LCs and analyzed how DNA changes are generated and accumulate in LCs. **I.** The DNA changes that contribute to cell transformation of a LC can be generated in the LCs in bone marrow, thymus, lymph nodes (LNs), and/or lymphoid tissues (LTs). In LNs and LTs, repeated pathogen-infections may be the main cause for cell injuries and DNA injuries of LCs. In marrow cavity, repeated bone-remodeling during bone-growth and bone-repair, by producing toxic substances, may be a source of damage to hematopoietic cells, including hematopoietic stem cells (HSCs) and developing LCs. In thymus, thymic involution and death of stromal cells may be a damaging factor for the developing T-cells. **II.** Point DNA mutation (PDM) and chromosome change (CC) are the two major types of DNA changes. CCs include numerical CCs (NCCs) and structural CCs (SCCs). Generation of PDM is often a result of Misrepair of DNA on a double-strand DNA break. Generation of NCC is rather a consequence of dysfunction of cell division promoted by damage. Generation of SCC may be a result of Misrepair of DNA on multiple DNA breaks. DNA breaks in a LC can be introduced not only by external damage, but also possibly by DNA rearrangements of *Ig/TCR* genes. **III.** Repeated cell injuries and repeated cell proliferation drive accumulation of DNA changes in LCs and HSCs. However, long-term accumulation of DNA changes occurs mainly in long-living stem cells including HSCs and memory cells. **In conclusion**, the DNA changes in LCs are generated and accumulate as a consequence of repeated cell injuries and repeated cell proliferation; and three potential sources of cell injuries of LCs may be: repeated bone-remodeling, long-term thymic involution, and repeated pathogen-infections.

**Keywords**

Lymphoid leukemia (LL), lymphoma, lymphoid cells (LCs), DNA changes, bone-remodeling, marrow cavity, thymic involution, pathogen infections, lymph nodes (LNs), point DNA mutation (PDM), numerical chromosome change (NCC), structure chromosome change (SCC), Misrepair of DNA, repeated cell injuries, and accumulation of DNA changes

## I. Introduction

Lymphoid leukemia (LL) and lymphoma are two large groups of blood cancers developed from lymphoid cells (LCs). Differently from most solid tumors, blood cancers can occur in young children. Some forms of leukemia and lymphoma such as acute lymphoblastic leukemia (ALL) and Burkitt lymphoma (BL) have higher incidences in children than in adults. For understanding pediatric LL and lymphoma, we have made a profound analysis on the causes and the mechanism of transformation of a LC. Our analysis was driven by the following questions:

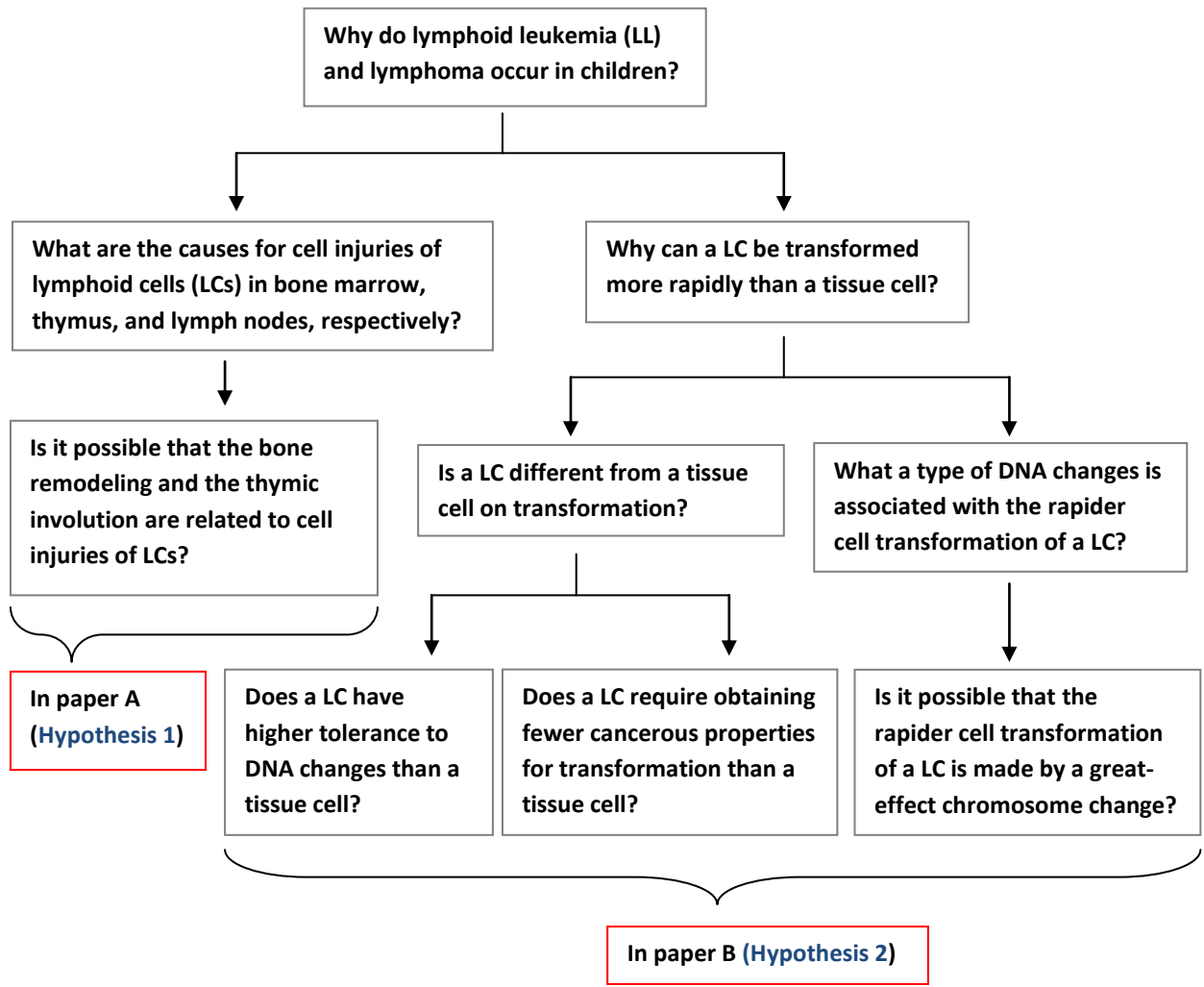

We will introduce our analysis and our hypotheses in four papers, namely: **paper A** (Three potential sources of cell injuries of lymphoid cells associated with development of lymphoid leukemia and lymphoma (the present paper)), **paper B** (Three pathways of cell transformation of lymphoid cell: a slow, a rapid, and an accelerated), **paper C** (Acute lymphoblastic leukemia may develop as a result of rapid transformation of a lymphoblast triggered by repeated bone-remodeling during bone-growth), and **paper D** (Pediatric lymphoma may develop by "one-step" cell transformation of a lymphoid cell).

In the present paper, we discuss the potential sources of cell injuries of LCs and analyze how DNA changes are generated and accumulate in LCs. In general, LL arises from a LC transformed in bone marrow, and lymphoma arises from a LC transformed in a lymph node (LN) or a lymphoid tissue (LT). However, because of the recirculation of LCs in bloodstream and lymph, some of the DNA changes that contribute to lymphoma development may be generated in precursor HSCs or LCs in marrow; and some DNA changes that contribute to LL development may be generated in a LC in a LN/LT. Therefore, it is useful to study LL and lymphoma together on their causes and their developing mechanisms. "Lymphoid cells (LCs)" include all the cells of lymphoid lineage in marrow, bloodstream, LNs, and LTs. The term "lymphocytes" is referred to specifically the naïve lymphocytes (Steven, 2016). We aim to show by our discussion that, the DNA changes in LCs are generated and accumulate as a consequence of repeated cell injuries; and three potential sources of cell injuries of LCs are: repeated bone-remodeling, long-term thymic involution, and repeated pathogen-infections.

We use the following abbreviations in this paper:

ALCL: anaplastic large cell lymphoma
ALL: acute lymphoblastic leukemia
AML: acute myeloid leukemia
ATLL: adult T-cell lymphoma/leukemia
APC: antigen-presenting cell
BL: Burkitt lymphoma
CC: chromosomal change
CLL: chronic lymphocytic leukemia
DLBCL: diffuse large B-cell lymphoma
EBV: Epstein-Barr virus
FL: follicular lymphoma
GC: germinal center
HC: hematopoietic cells
HIV: human immunodeficiency virus
HLTV-1: human T-cell lymphotropic virus type 1

HSC: hematopoietic stem cell
LC: lymphoid cell
LN: lymph node
LT: lymphoid tissue
MALTL: mucosa-associated lymphoid tissue lymphoma
MC: myeloid cell
MCL: mantle cell lymphoma
MDS: myelodysplastic syndrome
NCC: numerical chromosome change
NHL: non-Hodgkin's lymphoma
PDM: point DNA mutation
SCC: structural chromosome change
TCR: T-cell receptor

## II. Development of lymphoid cells (LCs) in marrow cavity and in thymus

After birth, all of our blood cells including LCs and myeloid cell (MCs) are produced by the hematopoietic stem cells (HSCs) in marrow. Bone marrow is a spongy tissue stored in the marrow cavity and the spongy part of a bone. Hematopoietic cells in bone marrow are a mixture of cells of different lineages and at different developing stages. They include HSCs, progenitor cells, blast cells, pro-cytes, and mature blood cells (Figure 1). Every day, about $10^{11}$ of new red cells and $10^8$ of new lymphocytes are produced in a human. For producing such huge number of cells, hematopoietic system needs to work constantly. In any moment, a large number of hematopoietic cells are in processes of cell division and cell differentiation.

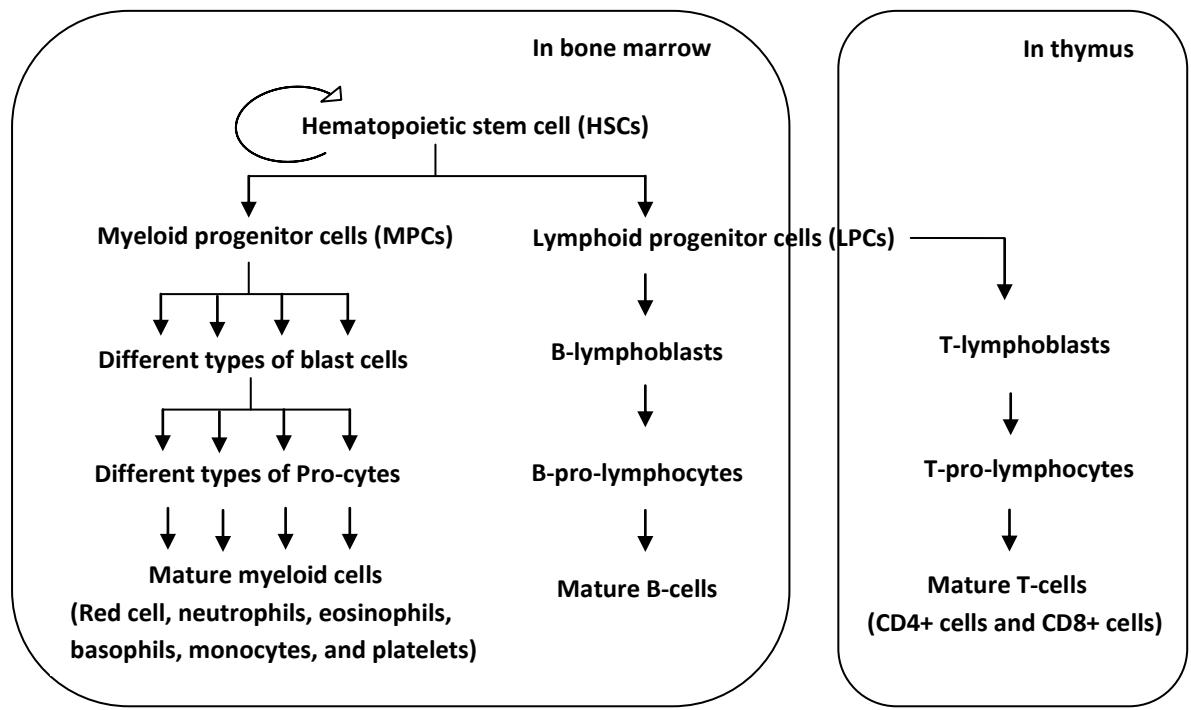

Figure 1. Development of lymphoid cells in bone marrow and in thymus

Hematopoietic stem cells (HSCs) are the stem cells for all types of blood cells. For hematopoiesis, a HSC needs to produce a progenitor cell by asymmetrical cell division. There are two types of progenitor cells: myeloid progenitor cells (MPCs) and lymphoid progenitor cells (LPCs). MPCs are the precursor cells for all the cells in myeloid lineage, including red cell, neutropjils, eosinophils, basophils, monocytes, and platelets. LPCs are the precursor cells for all the cells in lymphoid lineage, including T-cells, B-cells, and natural killer cells. Except T-cells, all types of blood cells are fully developed in bone marrow. T-cells develop in thymus from a LPC produced in marrow but migrated to thymus. Once produced, the progenitor cells proliferate rapidly and differentiate into blast cells in different lineages, such as myeloblasts, monoblasts, and lymphoblasts. A blast cell continues to proliferate and differentiate into one or more types of pro-cytes, such as pro-myelocytes (neutrophilic, basophilic, or eosinophilic), pro-monocytes, and pro-lymphocytes. A pro-cyte divides several times and produces mature cells of the same type.

## 2.1 Hematopoietic stem cells (HSCs): multi-potent and self-renewable

HSCs are the ancestors for all hematopoietic cells and all blood cells. A general process of hematopoiesis is as follows: **A.** a HSC produces a progenitor cell; **B.** the progenitor cell then proliferates rapidly to produce a large number of blast cells in different lineages; **C.** a blast cell then proliferates and differentiates into different groups of pro-cytes (immature cells); and **D.** a pro-cyte then divides for several times to become mature cells (Figure 1). HSCs account for only 0.001% of hematopoietic cells, but they can regenerate for maintaining their number. A HSC has the potential of self-renewal for our whole lifetime. By symmetrical cell division, a HSC produces two daughter HSCs. By asymmetrical cell division, a HSC gives rise to two different cells: a new HSC and a progenitor cell. Two major types of progenitor cells are myeloid progenitor cells (MPCs) and lymphoid progenitor cells (LPCs). MPCs are the ancestors for all the cells in myeloid lineage, including granulocytes (neutrophiles, basophiles, and eosinophiles), monocytes, erythrocytes (red cells), and megakaryocytes (platelets). LPCs are the ancestors for all the cells in lymphoid lineage, including T-cells, B-cells, and natural killer (NK) cells. In a normal marrow, 55% of nucleated cells are of granular lineage, 20% are of erythrocytic lineage, and 20% are LCs.

Since HSCs are the unique source of new blood cells, an abnormality on the number or on the functionality of HSCs may lead to a severe disease. For example, if regeneration of HSCs is disturbed by damage or by a genetic disorder, blood cells cannot be sufficiently produced, and aplastic anemia may occur. If a HSC is transformed by DNA changes, accumulation of abnormal HSCs may affect hematopoiesis. It is reported that the background DNA mutations in HSCs are related to the clonal hematopoiesis in old people (Shlush, 2015). HSCs are multi-potent and regenerable for our whole lifetime. Therefore, transplantation of autologous or allogeneic HSCs is an important replacing treatment in some malignant diseases, including leukemia, aplastic anemia, and myelodysplastic syndrome (MDS).

**2.2 Progenitor cells and lymphoblasts: oligo-potent, proliferative, and immature on cell functions**

Progenitor cell is the ancestor cell for one lineage of cells. For example, MPC is for myeloid lineage and LPC for is for lymphoid lineage. MPCs and LPCs are also self-regenerable, but only for some generations. Once produced, a progenitor cell proliferates and differentiates into different types of blast cells. For example, a MPC differentiates into megakaryoblasts, erythroblasts, myeloblasts, and monoblasts. A LPC differentiates into T-lymphoblasts and B-lymphoblasts. A blast cell then proliferates and differentiates into one or more types of pro-cytes. For example, a myeloblast will differentiate into three types of pro-myelocytes: neutrophilic, basophilic, or eosinophilic (Figure 1). A lymphoblast will differentiate into pro-lymphocytes. Lymphoblasts can be identified by specific cell surface markers such as CD10 and TdT. Lymphoblasts express intracellular IgM but not membrane IgM (mIgM). Except T-cells, all types of blood cells are fully developed in marrow. T-cells develop in thymus, from a LPC produced in marrow but migrated to thymus.

Progenitor cells and blast cells have three characteristics: oligo-potent, proliferative, and immature on cell functions. Thus, these cells have a risk of DNA injuries during DNA synthesis and cell division by damaging factors. If cell differentiation of one of these cells is altered by DNA changes, hematopoiesis can be severely affected. For example, when cell differentiation of a MPC is altered, MDS may occur. Failure of cell differentiation of a lymphoblast may result in occurrence of ALL.

## 2.3 DNA rearrangement of *Ig/TCR* genes in lymphoblasts

Lymphoblasts are the precursors of pro-lymphocytes and naïve lymphocytes. A naïve lymphocyte expresses membrane immunoglobulin (Ig) (on B-cell) or T-cell receptor (TCR, on T-cell). However, the molecules of Ig or TCR on different lymphocytes are different by the ligand (antigen)-binding part. This is a result of DNA rearrangement of *Ig/TCR* genes during differentiation of lymphoblasts (Chowdhury, 2004; Thomas, 2009). In human B-cells, an Ig protein has two identical heavy chains and two identical light chains. Each chain has a variable region. The variable region is coded by a gene composed of two or three of the following DNA segments: variable (V) segment, joining (J) segment, and divers (D) segment. A gene for the variable region of heavy chain is composed by V-D-J, and that for light chain is made by V-J. In fact, in locus of *Ig* genes (14q32) in B-lymphoblasts, there are many candidate segments for $V_H$, $D_H$, and $J_H$, respectively. For example, in human being, there are 51 (65) $V_H$-segments, 6 $J_H$-segments, and 27 $D_H$-segments for heavy chain, and there are 40 (70?) $V_L$-segments and 5 $J_L$-segments for light chain (Berg, 2002; Janeway, 2001; Jung, 2006).

DNA rearrangement of *Ig* gene is a process of recombining randomly one of $V_H$-segments, one of $J_H$-segments, and one of $D_H$-segments in locus of *Ig* gene, and cutting off other unneeded segments. By this process, each cell will have only one gene (V-D-J) for the variable region of heavy chain and one gene (V-J) for light chain. As a result of random recombination of V, J, and D segments, the final genes of *Ig* are different in different cells. The diversity of genes for the variable region of heavy chain is thus 51 ($V_H$) x 6 ($J_H$) x 27 ($D_H$) = 8262, and that for light chain is 40 ($V_L$) x 5 ($J_L$) = 200. An Ig molecule has two identical heavy chains and two identical light chains, thus an Ig has totally two different variable regions. The diversity of Ig is thus 51 ($V_H$) x 6 ($J_H$) x 27 ($D_H$) x 40 ($V_L$) x 5 ($J_L$) = 1652400. Probably, the Ig molecules produced by different naïve lymphocytes are distinct from each other on the antigen-binding part. DNA rearrangement of *Ig* gene increases largely the diversity of Ig molecules of B-cells. A large diversity of Ig molecules permits B-cells to recognize antigens in large diversity and produce antibodies in large diversity.

## 2.4 Pro-lymphocytes: uni-potent, proliferative, in large number, and immature on cell functions

Pro-cytes have higher degree of cell differentiation and maturation than blast cells, but they are not yet fully developed. Pro-cytes are proliferative but uni-potent (Figure 1). Each pro-cyte will differentiate into a number of mature cells of the same type. For example, a pro-lymphocyte develops into a number of mature (naïve) lymphocytes. Pro-lymphocytes can produce membrane Ig (mIg). Pro-lymphocytes can be distinguished from lymphoblasts and mature lymphocytes by specific surface markers including CD19, MCH (II), and CD20. CD10 and TdT are not expressed on pro-lymphocytes. The total number of pro-cytes may be much higher than that of blast cells, because a blast cell needs to divide several times to develop into pro-cytes. Thus pro-cytes may account for the largest part of immature nucleated cells in marrow. Since pro-lymphocytes are proliferative, they have also high risk of cell injuries during DNA synthesis and cell division. ALL and AML may arise also from pro-cytes, respectively from pro-lymphocyte and pro-myelocyte.

## 2.5 Naive lymphocytes: small, non-proliferative, and mature on cell functions

For distinguishing the new mature lymphocytes from the activated mature LCs (immune cells), we call the new-born mature LCs as "naïve lymphocytes". B-cells develop fully in marrow whereas T-cells develop in thymus. Naive lymphocytes will enter bloodstream and then lymph system. Part of them will be stored in spleen and LNs. Naïve lymphocytes are characterized by expression of mIgM and mIgD. Being non-proliferative, naive lymphocytes may have lower risk of DNA injuries and DNA mutations than lymphoblasts and pro-lymphocytes. Naïve lymphocytes are the smallest cells in our body. They have normal-sized nucleus but little cytoplasm. This suggests that a naïve lymphocyte produces much fewer intracellular proteins than other cells.

## 2.6 Development of T-cells in thymus

Thymus is an organ located anterior and superior the mediastinum between two lungs. This organ is made by two types of tissues: thymic stroma and hematopoietic cells. The hematopoietic cells are mainly developing T-cells, called also thymocytes. Thymic stromal cells include epithelial cells and dendritic cells. Stromal cells are important in the development of T-cells, because they provide a special microenvironment for selecting functional and self-tolerant T-cells (Nitta, 2016). Thymus has the largest size in late childhood (age 11-13), but it starts to shrink since age of puberty. Atrophy of thymus occurs mainly to thymic stroma, in which stromal cells are gradually replaced by fat tissues. Finally in an adult, thymus is made by some tiny islands hidden in fat tissues. Despite of thymic involution, T-cell hematopoiesis continues throughout our whole lifetime.

Thymus has two parts of structures: the cortex and the medulla. The cortex is the location for early events of T-cell development, including *TCR* gene rearrangement and positive selection of T-cells. The medulla is the location for later events of T-cell development, including negative selection of T-cells. Positive selection is to select the functional T-lymphoblasts, namely the lymphoblasts that can recognize self-MHC (major histocompatibility complex) presented by epithelial cells. Negative selection is for removing the auto-reactive T-cells, namely the T-cells that interact with the peptides presented by thymic dendritic cells. On cell functions, mature T-cells have two groups: CD4+ T-cells and CD8+ T-cells. CD4+ cells and CD8+ cells will be activated in LNs/LTs to develop respectively into T-helper cells (Th cells) and cell-toxic T-lymphocytes (CTLs).

TCR determines the antigen specificity of a T-cell. A large diversity of TCR enables T-cells to recognize antigens in large diversity. A TCR molecule has two chains: α chain and β chain. In some cases, TCR is composed by γ chain and δ chain. Similar to that in B-cells, DNA rearrangements of genes of α chain (locus: 14q11.2) and β chain (locus: 7q34) of TCR occur in T-lymphoblasts (Thomas, 2009; Janeway, 2001). The variable region of each chain is coded by a gene composed of two or three DNA segments (V-, D-, and J-segments): V-J for α chain and V-D-J for β chain. By DNA rearrangement of *TCR* gene, different T-cells will express different TCRs, which have different affinities to an antigen.

## III. Activation of naïve lymphocytes by pathogens in a lymph node (LN) or lymphoid tissue (LT)

Naïve lymphocytes will enter lymphatic system to be activated and then transported to all organs. Lymphatic system is part of immune system distributing all over our body. This system is composed of lymphatic organs, lymph vessels, lymph, and LCs. In human being, lymphatic organs include bone marrow, thymus, spleen, tonsil, LNs, and low-concentrated LTs. Lymph vessels are the tubes for connecting lymphatic organs and conducting lymph fluid from LN to LN and finally into bloodstream. Lymphatic system plays three roles in immunity: trapping and removing pathogens in LNs; activating lymphocytes by pathogens in a LN/LT, and conducting lymphocyte-recirculation in bloodstream and lymph.

### 3.1 Structure and functions of a LN

About 600 LNs distribute throughout our body. A LN has two parts of structures: the outer cortex and the inner medulla. Naïve B-cells are stored in the superficial part of cortex and mostly in the mantle zones of follicles. Activation of a B-cell takes place inside a follicle in deep part of cortex. T-cells are stored and activated in center of paracortical zone. B-cells and T-cells can be distinguished by some cell markers: CD20, CD79a, and PAX5 are specific for B-cells; and CD5 is specific for T-cells. The LCs in bloodstream enter a LN via high endothelial venues. A LN has two functions: **A.** activation of naïve lymphocytes by pathogens and production of antigen-specific immune cells; and **B.** filtration and clearance of pathogens and foreign substances in lymph. In a LN, there are not only B-cells and T-cells but also other types of immune cells including dendritic cells, granulocytes, macrophages, and NK cells. These innate immune cells can recognize pathogens and foreign substances in lymph and remove them. Dendritic cells are antigen-presenting cells (APCs), and they play an important role in activation of B-cells and T-cells.

Low-concentrated LT is a loose organization of lymphocytes in part of a tissue. It is composed by B-cells and T-cells, but it does not have fixed structure and capsule. In a LT, there are M-cells, which function as antigen-presenting cells. An M-cell can trap and present antigens to Th cells for activating B-cells and T-cells. LTs distribute widely in skin, mucosa of digestive tract and airway wall, and the capsules of an organ. In mucosal LT, T/B-cells are located in sub-epithelium, but M-cells are located between epithelial cells. In skin and capsules of organs, LTs distribute not only in sub-epithelium and but also around dermal veins (Ono, 2015; Nomura, 2014). The peri-vascular LCs may function as a barrier to prevent the invasion of pathogens from infected tissues into bloodstream and that from bloodstream into tissues.

### 3.2 Activation and differentiation of B-cells into immune cells in a LN

Activation of a naïve lymphocyte by pathogens is a complex process. Here we give only a brief introduction of this process. Activation of a B-cell takes place in a follicle in superficial cortex, and activation of a T-cell occurs in the paracortical zone. The activation of a naive lymphocyte by the primary contact with an antigen presented by APC is called primary immune response. By primary response, a naïve lymphocyte differentiates into a large number of antigen-specific immune cells, including effector cells and memory cells. Plasma cells are effector B-cells, which can produce antigen-specific antibodies in LNs and in affected tissues. Th cells and CTLs are effector T-cells. Effector cells and memory cells will be transported to all organs via lymph and bloodstream to take part in adaptive immunoreactions in infected tissues. When a memory cell contacts again the same type of antigen, it will be activated quickly and produce again a large number of effector cells and memory cells. This process is called secondary immune response.

Activation of B-cell needs two signals: binding of B-cell receptor (mIg) to an antigen presented by APC and interacting of the B-cell with a Th cell. Upon antigen-activation, B-cells proliferate quickly and the produced cells compose a germinal center (GC) in a follicle. The size of GC is enlarged gradually with the increasing of number of cells. Figure 2 shows a general process of activation and differentiation of a B-cell in a LN. In primary response, a B-cell proliferates and differentiates firstly into centroblasts. A centroblast then proliferates and differentiates into a large number of centrocytes. Centrocytes are then selected by antigen-affinity through interacting with two cells: follicular antigen-presenting dendrite cells and Th cells. The centrocyte that has the highest antigen-affinity (mIg-antigen affinity) will continue differentiation and finally develop into immunoblasts. Other centrocytes will undergo apoptosis (Figure 2). An immunoblast then differentiates into two types of cells: plasmablasts and memory B-cells. A plasmablast will proliferate and produce a large number of plasma cells.

During activation of B-cell, the B-cells in different differentiating stages have different morphologies and distribute in different areas of a GC (Janeway, 2001). For example, naïve B-cells are small cells distributing in the mantle zone of GC. Centroblasts are mediate-sized cells distributing in the dark zone of GC. Since centroblasts are proliferative, they are basophilic and have blue color in hematoxyline-eosin (HE) staining. Centrocytes are small cells located in center (light zone) of GC. Centroblasts can be distinguished from centrocytes also by the shape of nucleus: the former have non-cleaved nucleus whereas the latter have cleaved nucleus. Immunoblasts differentiate into plasmablasts and memory cells in the cap zone of GC. Immunoblasts are large round cells but plasmablasts are large egg-shaped cells (Figure 2). Memory B-cells are small cells distributing in the marginal zone of a follicle. Plasma cells are large egg-shaped cells stored in medullary cords.

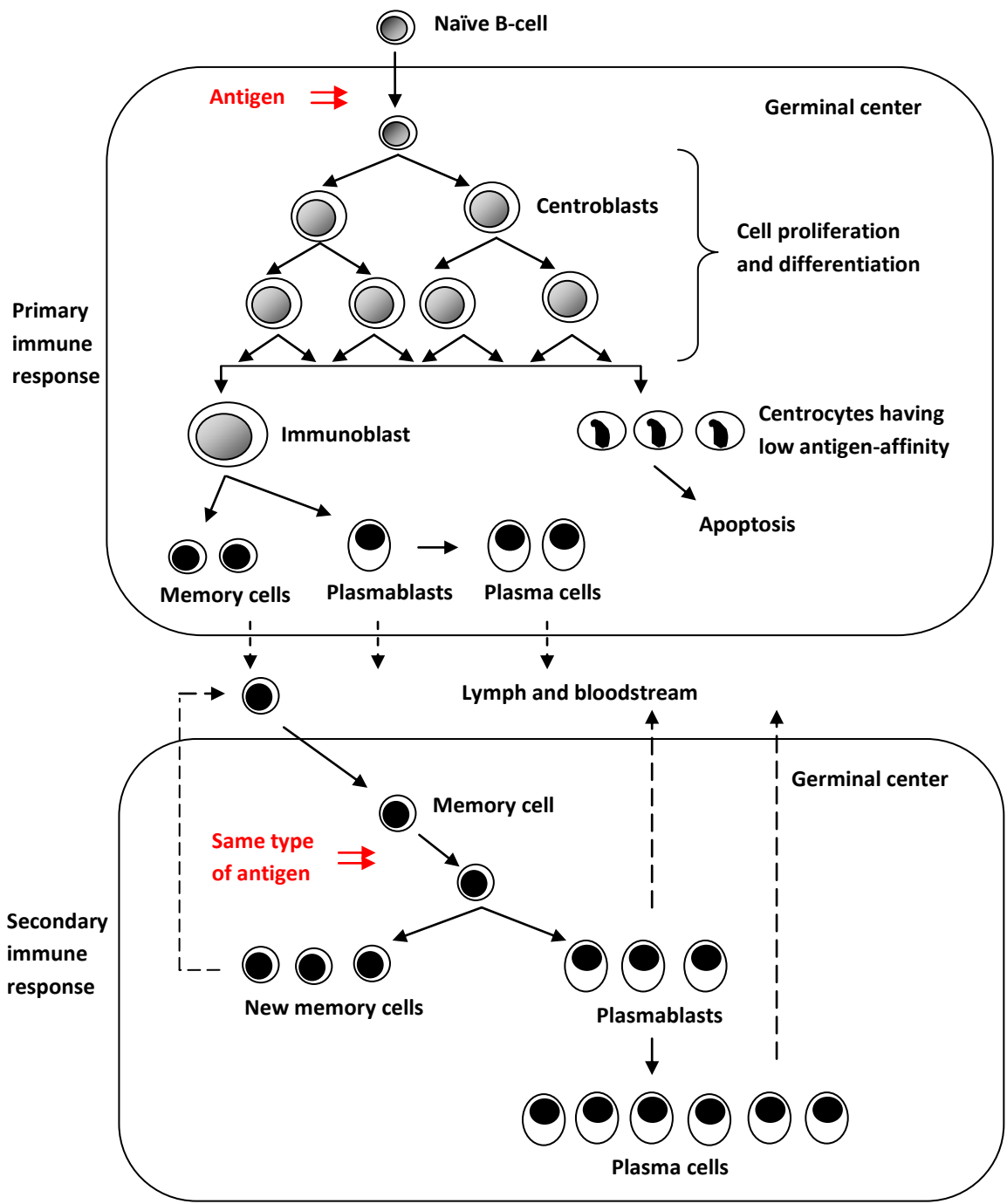

**Figure 2. Activation and differentiation of B-cells in primary and secondary immune responses**

Activation of a naive lymphocyte by primary contact with an antigen is called **primary immune response**. In primary response, a naïve lymphocyte proliferates and differentiates into a large number of antigen-specific immune cells, including effector cells and memory cells. When a memory cell contacts again the same type of antigen, the cell will be activated quickly and produce again a large number of effector cells and memory cells. This process is called **secondary immune response**. Activation of a B-cell takes place in a follicle. Upon activation by an antigen, a B-cell proliferates and differentiates firstly into centroblasts. A centroblast then proliferates and differentiates into a large number of centrocytes. The centrocyte that express the highest antigen-affinitive mIgs will differentiate into immunoblasts. Other centrocytes will undergo apoptosis. An immunoblast then differentiates into plasmablasts and memory B-cells. Memory B-cells and plasmablasts will be transported to all organs via lymph and bloodstream. A plasmablast can proliferate and produce a large number of plasma cells, which will produce antibodies.

### 3.3 Somatic hypermutation and class-switching recombination of *Ig* genes in B-cells

During activation of a B-cell, somatic hypermutation of *Ig* genes takes place in centroblasts. Hypermutation occurs to the part of DNA that codes the variable region of heavy or light chain of Ig protein. This process is triggered by an enzyme, namely the activation-induced deaminase (AID) (Dudley, 2005). By deaminating, AID can transform a bases-C in a strand of DNA into a uracil (base-U). When a base-C is converted into base-U, the double strands of DNA are not any more complementary (base U- base G) at this point. Then, the "head" (the part for pairing) of base-U is removed, leaving the base-G in complementary strand suspending in non-matching state. In subsequent DNA replications, this non-matching point will be paired by any type of base: N (A, T, C, or G). Thus, by this process, a base-G in a mother cell is randomly replaced by A, T, C, or G in grand-children cells. In this way, the Igs expressed in different centrocytes will have different motifs in variable regions, by which the cells have different affinities to an antigen.

Ig molecules have five isotypes by structure, and they are IgM, IgE, IgG, IgD, and IgA. These isotypes have different Fc fragments in Ig heavy chain. Generation of an isotype of Ig is a result of class-switching recombination of *Ig-Fc* genes, which may occur in immunoblasts. In locus of *IgH* genes, there are eight candidate C-segments for making *Fc* genes: α, μ, ε, δ, $\gamma^1$ $\gamma^{2a}$ $\gamma^{2b}$, and $\gamma^3$ (segment). Recombination of some of these segments and deletion of unneeded segments is essential for constructing a final *Fc* gene in a cell (Davila, 2001; Dadi, 2009). Different ways of recombination will result in generation of different forms of *Fc* genes. For example, the *Fc* gene for IgM has all of the eight C-segments: α, μ, ε, δ, $\gamma^1$ $\gamma^{2a}$ $\gamma^{2b}$, and $\gamma^3$; that for IgG have six C-segments: α, δ, $\gamma^1$ $\gamma^{2a}$ $\gamma^{2b}$, and $\gamma^3$; and that for IgA has only one α C-segment. As a result, different immunoblasts may produce different isotypes of Igs, although they may have the same antigen-specificity.

### 3.4 Activation and differentiation of T-cells into immune cells in a LN

Activation of a naïve T-cell needs two signals presented on an APC: antigen-MHC complex and B7 proteins (CD80 and CD86). The first signal is recognized by TCR, namely MCH-II-bound antigen is recognized by the TCR of CD4+ cells whereas MCH-I-bound antigen is recognized by the TCR of CD8+ cells. The second signal is recognized by CD28 on T-cells (Janeway, 2001). These combined signals then promote proliferation and differentiation of the T-cell. Finally, CD4+ T-cells differentiate into two types of effector cells and memory cells: antigen-specific regulatory T-cells (Treg cells) and T-helper cells (Th cells). CD8+ cells differentiate into antigen-specific cell-toxic T-lymphocytes (CTLs) and memory cells. Differently from that in B-cells, T-cells do not undergo somatic hypermutation during differentiation. Thus, a T-cell has less specificity and affinity to an antigen than a B-cell. In a LN, the T-cells at different differentiating stages may all distribute in the paracortical zone. Probably, except T-immunoblast (?), all T-cells, including naïve cells, effector cells (Treg cells, Th cells, and CTLs), and memory cells, are small cells.

## 3.5 Recirculation of lymphoid cells and homing tendency of memory cells

LCs at different developing stages have different lifespans. A naïve lymphocyte and an effector cell can survive for only several days and a plasmablast may have a lifespan of several weeks. Differently, a memory cell can survive for several years. Importantly, memory cells are regenerable for our whole lifetime. By each time of secondary immune response, a memory cell can be re-stimulated to produce new effector cells and new memory cells. New memory B-cells may be produced also in a GC, where the cells can be again selected by antigen-affinity (Figure 2). Via lymph, some LCs, including effector cells, memory cells, and un-activated naïve lymphocytes, can reenter bloodstream and then redistribute in all LNs and LTs. This is called lymphocyte-recirculation. A complete recirculation from a LN to the same LN takes 18-20 hours. By this recirculation, a LC can pass different organs and pass the same lymph organ many times. Lymphocyte-recirculation increases largely the opportunity for a lymphocyte to encounter an antigen.

Although transported to all organs, memory cells have a tendency to return to the locations where they are activated by antigens (Gregor, 2017; Shin, 2017). The homing behavior of memory cells is mediated by homing receptors, which can recognize the addrssin molecules on endothelial cells (Mueller, 2013; Roy, 2002). These homing receptors include L-selectin, CLA, LFA-1, VLA-4, CD4, and LPAM-2. They can bind respectively to the molecules of PNAD, E-selectin, ICAM-1, ICAM-2, VCAM-1, Mad, and CAM on endothelial cells. Due to this homing tendency, memory cells can be repeatedly reactivated by the same type of pathogens in infected tissues.

## IV. Repeated bone-remodeling during bone-growth and bone-repair may produce damage to marrow cells

DNA changes are the trigger for cell transformation of a LC into a cancer cell. However, generation of DNA changes is a consequence of cell injuries and DNA injuries. Exploring the causes for cell injuries of LCs is thus critical for understanding leukemia and lymphoma. Exposure to radiation or chemicals such as that in radiotherapy and chemotherapy may be a causing factor for cell injuries of LCs. However, most patients with leukemia/lymphoma do not have experiences of radiotherapy, chemotherapy, or contacting with pesticides. We find out that, apart from unspecific environmental factors, LCs may have three "specific sources" of cell injuries: repeated bone-remodeling during bone-growth and bone-repair, for the HSCs and B-LCs in marrow; long-term thymic involution, for the developing T-cells in thymus; and repeated pathogen-infections, for the LCs in LNs/LTs. The third source is known, but the first two are our hypotheses. We will discuss the roles of these three factors in cell injuries of LCs respectively in this and next two parts.

B-cells are produced in bone marrow and T-cells are produced in thymus. Thus, the cell injuries of LCs that are associated with development of lymphoid leukemia should be mainly produced in marrow cavity. ALL occurs mostly before age 25, in the period of time when bones are growing. This phenomenon suggests that bone-growth may be related to ALL development. It is known that bone-growth is a result of repeated modeling-remodeling of bone tissues. Thus, it is quite possible that the repeated bone-remodeling during bone-growth produces occasionally damage to marrow cells.

## 4.1 Formation of marrow cavity and spongy bone is a result of repeated modeling-remodeling of bone tissues during bone-development

Marrow cavity is the central cavity of a long tubular bone. The cavity is surrounded by contact bone as a long wall and by spongy bone at two ends. Spongy bone exists in the center of flat bones and in the two metaphysic parts of a long bone. Spongy bone is made by dense trabeculae. In a long bone, the lumen of marrow cavity and the tunnels of spongy bone are connected. Hematopoietic cells are produced in marrow cavity and the tunnels of spongy bone (Figure 3). Before age 40, most bones can produce blood cells. However, since age 40, blood cells are mainly produced in the spongy part of flat bones. It is observed that leukemia cells can invade into bone tissues in late stage of leukemia. Thus, it is quite possible that a change of structure or function of bone tissues affect hematopoietic cells.

Marrow cavity and spongy bone are developed and enlarged with the growth of a bone. Bone development is achieved by two parallel processes: formation of new bone matrix (modeling of bone) by osteoblasts and absorption of unused bone matrix (remodeling of bone) by osteoclasts for reshaping the bone (Horton, 1990). Osteoblasts are produced by periosteum, but osteoclasts are produced by hematopoietic cells in marrow. A bone develops from a cartilage model. Ossification of the cartilage model starts from two ossification centers. The primary ossification center is located in center of the model, and it starts to produce bone tissues before birth of baby. The second center is located at two ends of the model and starts to work after birth of baby. Marrow cavity is developed from the primary ossification center by several steps. Firstly, new cartilage tissues need to be continuously produced from ossification center for enlarging the size of the cartilage model. New cartilage tissues are then modeled into bone tissues through adding bone matrix by osteoblasts. The outer parts of new bone tissues are then repeatedly modeled and remodeled to develop into compact bone. In the same time, the inner parts of new cartilage/bone tissues are gradually absorbed (remodeled) by osteoclasts, to develop into a cavity**.**

Spongy bone is produced by the secondary ossification center of a long bone. Formation of spongy bone is a result of partial digestion of new cartilage/bone tissues by osteoclasts in the metaphyses of a bone. However, ossification of different bones starts at diffident ages. Ossification of big bones starts before birth, but that of small bones including flat bones and small long bones starts mostly after birth. However, all bones will have ossification centers before age 10. This suggests that the period of age 0-10 may be a peak time of formation of new marrow cavities and spongy bones.

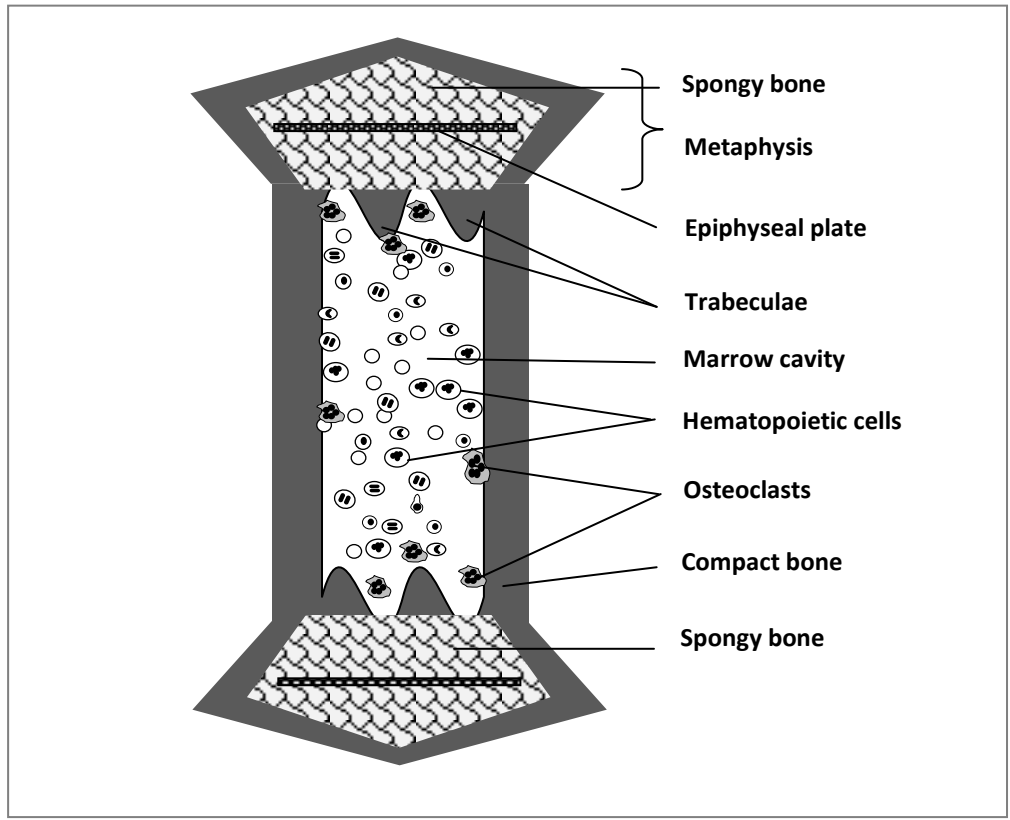

**Figure 3. Blood cells are produced in marrow cavity and the spongy part of a bone (a schematic graph)**

Hematopoiesis proceeds mainly in marrow cavity and the spongy part of a bone. Marrow cavity is the central cavity of a long tubular bone. The cavity is surrounded by contact bone as a long wall and by spongy bone at two ends. Spongy bone exists in the center of flat bones and in the two metaphyses of a long bone. Spongy bone is made by dense trabeculae. In a long bone, the lumen of marrow cavity and the tunnels of spongy bone are connected. Blood cells are produced in marrow cavity and the tunnels of spongy bone. During bone development, repeated remodeling of bone tissues is needed for reshaping the bone. Bone-remodeling of occurs to the part of bone exposed to cavity or tunnels. Bone-remodeling is a result of absorption of bone tissues by osteoclasts in marrow. Osteoclasts are produced by hematopoietic cells.

Marrow cavity and spongy bone will be enlarged with the growth of a bone. A long bone grows in length and in thickness in the same time. Growth in length is a result of repeated additions of new cartilage tissues to metaphyses from epiphyseal plates. The outer part of new cartilage tissue is then modeled-remodeled into compact bone. The inner part is absorbed by osteoclasts and becomes a part of cavity. In this way, the cavity is enlarged gradually in length, and the lumen of spongy bone and that of marrow cavity join at metaphysis. Increasing of diameter of a long bone is a result of two processes: addition of new bone tissues beneath periosteum and absorption (remodeling) of old bone tissues that are exposed to cavity. In this way, the thickness of the bone is increased and the diameter of the cavity is as well enlarged. In the meantime, the spongy bone in metaphysis is also remodeled, reshaped, and enlarged. Importantly, for bone-growth, the process of modeling-remodeling of bone tissues is repeating all the time in a child, for more than 20 years!

**4.2 Bone-remodeling occurs also during bone-repair and bone-adaptation**

In fact, bone-remodeling takes place also after bone maturation. To give a bone an adaptable resistance to external load, bone-remodeling may occur often to the compact bone and/or the spongy part of a bone. This adaptive bone-remodeling increases the potential of functionality of a bone and reduces the risk of bone fracture (Horton, 1990). However, bone injuries will occur anyway when a load exceeds the limit. Apart from visible fractures, a bone may have occasionally tiny injuries. In situations of bone injuries, bone-repair will be promoted for re-linking bone tissue. Bone-repair is also a process of modeling-remodeling of bone tissues to rebuild the structure of part of a bone. A bone will have increased risk of injuries when it becomes osteoporotic and fragile at old age. Bone injuries and bone-repair have thus increased occurrences in old people. Therefore, bone-remodeling may have the highest frequency in two periods of time in our life: at young age (age 0-25) when the bones are growing; and at old age (age over 60) when the bones become fragile.

**4.3 Repeated bone-remodeling may produce toxic substances to hematopoietic cells in marrow**

Bone-remodeling is an essential process in bone-development and bone-repair. Remodeling of bone tissue occurs to the part of bone exposed to cavity or tunnel. Therefore, digestion of bone matrix by osteoclasts during bone-remodeling may produce damage to hematopoietic cells. An osteoclast absorbs bone tissue in three steps: **A.** the osteoclast adhere to the surface of the bone exposed to cavity (or tunnel); **B.** the osteoclast produce acid substances and enzymes including matrix metalloproteinases (MMPs) to digest local bone matrix; and **C.** the osteoclast enwrap and swallow the wastes of digested bone (Väänänen, 2000). Normally, this process is accurately controlled and may not affect neighbor cells. However, if this process is disturbed somehow, the digestive substances may injure local hematopoietic cells (HCs). It is known that the children at age 2-5 have high incidence of bone fractures due to their fragile muscles and bones. Notably, a bone fracture or micro bone injury in a child may perturb the normal bone-remodeling for bone-growth. Thus, a child may have higher risk of injuries of HCs by (disturbed) bone-remodeling than other ages. In addition, for most small bones and flat bones, formation of marrow cavity occurs at age 0-10. This factor may also increase the risk of injuries of HCs by bone-remodeling in a child. There is a layer of fat tissue between bone tissue and bone marrow, and this fat tissue can protect HCs as a barrier. Nevertheless, the cells beneath bone trabeculae may have a risk to be injured by the aggressive substances produced during bone-digestion.

**V. Long-term thymic involution may produce damage to the developing T-cells in thymus**

Thymus is an organ for T-cell development. However, this organ undergoes involution since age of puberty. The weight of thymus is 15 g at birth and 35 g at age 11-13. Since puberty, thymus starts to shrink by death of stromal cells. Finally, at age 50, most part of thymus is replaced by fatty tissues. It is known that T-lymphoblastic lymphoma/leukemia (T-LBL) originates from T-lymphoblast. 60%-80% of T-LBL patients have expanded mediastinal mass, indicating that the tumor of T-LBL may start from thymus. T-LBL occurs mainly in adolescents and young adults, but rarely in old people. This suggests that the cell transformation of T-lymphoblast in T-LBL may be related to thymic involution. During thymic involution, death of stromal cells may release a great deal of enzymes and toxic substances. These substances, if not removed in time, may injure the developing T-cells (thymocytes). Although the risk of injury of T-cells in such a physical process is rare, long-term and constant thymic involution can largely increase this risk.

Taken together, repeated bone-remodeling and constant thymic involution may be two internal damaging factors for HCs, including HSCs, developing B-cells, and developing T-cells. Cell injuries by internal damage can occur also to other cells including tissue cells during body-development and during inflammations. DNA changes can be also generated in some of injured tissue cells. However, for a tissue cell, cell transformation occurs mainly at old age, as a co-effect of external and internal damaging factors. Differently, a LC can be transformed at young age as that seen in ALL. Thus, the effect of internal damage on a LC can be recognized.

## VI. Repeated infections are the main cause for the cell injuries of LCs in LNs and LTs

In lymphoma development, cell transformation of a LC is often an effect of multiple driver DNA changes. However, the last driver DNA change should be generated in the first transformed cell in a LN/LT. Exposures to toxic chemicals such as herbicides and pesticides may be a cause for DNA changes in LCs. However, for most patients who live in different environments, pathogen-infections should be more important in causing DNA changes in the LCs in LNs/LTs. Bacterial and viral can both damage LCs, but viral are more carcinogenetic, because viral can proliferate inside the host cells and damage the host DNAs directly.

**6.1 Three major types of pathogens associated with lymphoma development**

LNs are the locations where pathogens are trapped and filtered. The trapped pathogens in a LN can be removed by the high-concentrated immune cells, including effector LCs, granulocytes, macrophages, and NK cells. Our immune system has high efficiency on killing bacterial and extracellular viral, but it cannot remove all the viral that hide in a host cell. When an infection is severe or when our immunity is too low, excessive pathogens can damage the cells in LNs/LTs and in affected tissues. Bacterial can injure host cells by producing extoxin and endotoxin. Viral can damage a host cell via three pathways: adhesion to and lysing a host cell, self-multiplication in a host cell then lysing the cell, and integration of a viral DNA into a host DNA. Importantly, the viral hidden in a host cell can cut host DNAs on multiple points. Viral DNA can be inserted into a host DNA via DNA breaks.

Insertion of a viral DNA into a host DNA has two consequences for the host cell: DNA change and genome instability.

Three types of pathogens are closely associated with lymphoma development, and they are Epstein-Barr virus (EBV), human T-cell lymphotropic virus type-1 (HTLV-1), and *Helicobacte-Pylori* (*H. pylori*). Human immunodeficiency virus (HIV) is also related to lymphoma development. A patient with acquired immunodeficiency syndrome (AIDS) has high risk on developing lymphoma. However, the role of HIV in AIDS-related lymphoma is indirect, not by making DNA changes in LCs but rather by impairing T-cell-mediated immunity (Krishnan, 2014). In fact, different types of pathogens invade our body via different routes. The starting site of a lymphoma is determined by the type of pathogen.

EBV is a type of DNA virus, and it spreads widely in the world. Most EBV-affected people are asymptomatic EBV-carriers (Geng, 2015). EBV has high affinity to B-LCs and nasopharyngeal epithelial cells. EBV is known to be associated with development of B-cell lymphoma, including Burkitt lymphoma (BL), diffuse large B-cell lymphoma (DLBCL), and Hodgkin's lymphoma (HL) (Grywalska, 2015). EBV can hide in nasopharyngeal tissue and be transmitted via saliva. Mouth to mouth kiss may be an important transmitting route of EBV for human beings (Dunmire, 2015). Infectious mononucleosis (IM) is a diseases caused by EBV infection. The kissing-habit in western countries may be related to the higher incidence of IM in young adults in these countries (Womack, 2015). EBV infects firstly the mucosa of digestive tract and that of airway. The neck LNs are responsible for collecting the lymph from mucosa of mouth and airway. Thus, in EBV infections, the LCs in neck LNs and that in mucosal LTs will be firstly affected by EBV. This explains why B-cell lymphomas start often from mucosa (in BL) or from the LNs in the neck-supraclavicular area (in HL).

HTLV-1 is a type of retrovirus that has high infinity to mature CD4+ T-cells. Infection of HTLV-1 takes place only in some areas of the world, including Japan, Caribbean, South America, Iran, and Africa. HTLV-1 survives in bloodstream and tissue fluid in infected individuals. HTLV-1 is transmitted via sexual contact, breastfeeding, and blood transfusion. Although HTLV-1 infection is high in the above regions, only 5% of infected patients develop adult T-cell lymphoma/leukemia (ATLL) at adult age (Oliveira, 2017).

*H. Pylori* are a type of bacteria that is associated with cancer development. *H. pylori* are a causing factor not only for stomach cancer but also for mucosa-associated lymphoid tissue lymphoma (MALTL) (Hu, 2016). Infection of *H. pylori* is common in all countries, and the infected patients have often chronic gastric ulcers. *H. pylori* can injure mucosal cells including epithelial cells and LCs. Long-term infection of *H. Pylori* is needed for MALTL development (Park, 2014). *H. pylori* eradication is an effective pre-treatment for MALTL.

**6.2 Immunodeficiency: as a main causing factor for repeated pathogen-infections**

The pathogens that are related to lymphoma development have all high infection rates in population. However, only a small group of infected patients develop lymphoma. For example, 90% of the people in the world are carriers of EBV, but the incidence rate of B-cell lymphoma is only 0.01% in whole population (Grywalska, 2015). Infection of HTLV-1 occurs in 10% of population in Japan, but only 5% of infected individuals develop ATLL. Hence, the patients who develop lymphoma may be those who have higher frequency of infections than others. It is rather the frequency of infections that is critical in lymphoma development. Except in big epidemic, repeated infections are often a consequence of short-term or long-term immunodeficiency. Therefore, immunodeficiency contributes to lymphoma development probably by causing repeated infections.

Immunity insufficiency of an individual can be caused by several factors. The most common factors are: **A**. insufficiency of nutrition in children and sick people; **B**. infection of pathogens which can weaken immune system, such as HIV and malarial parasites; **C**. immune suppression by medicaments such as that applied in autoimmune diseases or organ transplantations; and **D**. aging of organs in old people. Hypo-immunity can have severe consequences. When pathogens cannot be cleared completely, they can invade into deep tissues even into bloodstream. Infections will occur repeatedly and chronically. The pathogens in large numbers can damage repeatedly LCs and tissue cells.

In USA, the overall incidence of lymphoma has slightly declined since 2012 (Teras, 2016). However, in the past 50 years before 2012, the incidence of non-Hodgkin's lymphomas (NHLs) was increasing in the world and in big cities (Skrabek, 2013). Several factors may be related to the increasing incidence of NHLs before 2012: increased populations in cities, increased usages of anti-inflammatory medicaments including aspirin, and increased rate of HIV-infection. Firstly, the people who live in big cities have increased risk of pathogen-infections thus have increased risk of lymphoma development. Secondly, people are now used to take anti-inflammatory drugs for anti-pains or anti-fever since a long time. These medicaments may unfortunately suppress our immunity.

Studies showed that long-term application of low-dose aspirin can reduce the risk of cancer development. Aspirin has this "anti-cancer" effect probably due to its role on inhibiting inflammation and tissue/cell repair. In our view, cell transformation in tumor-development is a result of accumulation of Misrepairs (mutations) of DNA (Wang-Michelitsch, 2015). Misrepair of DNA is a strategy of DNA repair when a cell has a severe DNA injury. However, aspirin can inhibit cell repair and DNA repair. Application of aspirin may result in failure of DNA repair (leading to cell death) rather than Misrepair of DNA (possible avoiding cell death). Thus for tissue cells, low-dose aspirin may slow down the accumulation of DNA mutations in cells and delay the occurrence of a cancer (Wang-Michelitsch, 2015). However, for LCs, aspirin may have an opposite effect. By inhibiting inflammations and immunoreactions, aspirin may increase the risk of cell injuries of LCs by pathogens. Therefore, aspirin may accelerate the accumulation of DNA changes in LCs and accelerate the occurrence of a lymphoma.

It is known that AIDS patients have high incidence of lymphomas. The global number of AIDS-related death was increasing each year before 2005. However, due to the invention of effective drugs, the number of HIV-infection starts to decrease since 1999 and that of AIDS-related death is reduced since 2006 (Roser, 2017). Thus, the incidence of HIV-related lymphomas may have declined in recent years, and this may contribute to the slight decline of incidence of NHLs since 2012.

In summary, apart from environmental factors, there may be three sources of damage that are more specific for LCs. They are: **A.** repeated bone-remodeling during bone-growth and bone-repair, for the HSCs/LCs/MCs in marrow; **B.** long-term thymic involution, for the developing T-cells in thymus; and **C.** repeated infections of viral or bacterial, for the T-LCs and B-LCs in LNs and LTs (Table 1). Bone-remodeling may be associated with ALL development. Thymic involution may be related to T-LBL development. BL and ATLL are rather caused by pathogen-infections.

**Table 1. Three potential sources of cell injuries of lymphoid cells**

| Locations | Potential sources of damage | Affected cells | Associated LL/lymphoma |
| --- | --- | --- | --- |
| **In marrow cavity** | Repeated bone-remodeling during bone-growth and bone–repair | HSCs, LCs, and MCs in marrow | **ALL** (acute lymphoblastic leukemia) |
| **In thymus** | Long-term thymic involution | Developing T-cells | **T-LBL** (T-lymphoblastic lymphoma/leukemia) ? |
| **In lymph nodes and lymphoid tissues** | Infections of viral or bacterial | B-LCs and T-LCs in LNs and LTs | **BL** (Burkitt lymphoma) and **ATLL** (adult T-cell lymphoma/leukemia |

## VII. DNA changes in LCs are generated often as results of Misrepair of DNA on DNA breaks

Modern sequencing techniques have made it possible to detect "all" the DNA changes in a cancer cell. However, only some of the DNA changes in a cancer cell contribute to cell transformation. Notably, in most forms of LL and lymphoma, the driver DNA changes in different patients are different. It would be too time-consuming and too expensive to study all the DNA changes in each patient. But one thing is clear: the more DNA changes a cell has, the higher risk has a cell on transformation. Thus, for understanding cancer development, it is more important to study how DNA changes are generated in a somatic cell and how they accumulate in cells.

### 7.1 Two types of DNA changes: point DNA mutation (PDM) and chromosome change (CC)

Point DNA mutation (PDM) and chromosome change (CC) are the two major types of DNA changes. A PDM exhibits as alteration, deletion, or insertion of one or two bases in a DNA. A PDM can affect at most "one" gene, thus it is also called gene mutation or molecular mutation. For example, *SOX11* mutation is found in 90% of mantle cell lymphoma (MCL) cases (Xu, 2010). Differently, a CC (also called cytogentic abnormality) can affect multiple genes. A CC can be numerical CC (NCC) or structural CC (SCC). A NCC exhibits as loss or gain of chromosomes of a cell. Major forms of NCCs include hyperdiploid, hypodiploid, aneuploid, loss of a chromosome (-), and trimosy (of chromosomes). For example, aneuploid is found in over 50% of ALLs (Al-Bahar, 2010; Braoudaki, 2012). Trimosy 8 is frequent in AMLs (Paulsson, 2007). A SCC exhibits as translocation (t), deletion (del), gain (+), or inversion (inv) of portion of a chromosome. For example, translocation t(8;14) is a main form of CC in Burkitt lymphoma (BL) (Cai, 2015), and t(2;5) is specific for ALK-positive anaplastic large cell lymphoma (ALK$^+$-ALCL) (Fielding, 2012). Notably, SCCs and NCCs are rare in cancers and sarcomas that arise from tissue cells, but not rare in leukemia and lymphoma. Different types of DNA changes are generated in cells by different mechanisms. Studies show that, PDM and SCC are generated as consequences of DNA breaks, but NCC is generated as a result of dysfunction of cell division.

**7.2 Generation of PDM: as a result of Misrepair of DNA on a double-strand DNA break**

On the mechanism of generation of PDM, there is still a debate. In our view, a PDM is neither an original DNA damage nor an error of DNA synthesis, but rather a result of Misrepair of DNA (Wang, 2009; Wang-Michelitsch, 2015). Studies have shown that Misrepair of DNA is the main source of PDMs (Natarajan, 1993; Bishay, 2001; Kasparek, 2011). A severe DNA injury including double-strand DNA break can promote SOS repair of DNA. One strategy of SOS repair of DNA is to re-link the broken DNA by non-homologous end joining (Rothkamm, 2002; Iliakis, 2015). Such SOS repair is essential for preventing failure of DNA and death of cell; however it may result in Misrepair of DNA. Misrepair of DNA is thus a strategy for cell survival. However, the surviving chance of a cell by Misrepair of DNA is low. There are three reasons for that: **A.** a cell that has DNA injuries may have also severe injuries on other parts of the cell; **B.** some DNA Misrepairs (mutations) are toxic to the cell; and **C.** a cell that is altered on phenotype by DNA mutations will be removed by immune system. Therefore, in situations of severe DNA injuries, most of injured cells will die, and only quite a few may survive by Misrepair of DNA. Thus, generation of PDM by Misrepair of DNA is a rare affair, and it is paid by death of many injured cells.

**7.3 Structural CC (SCC): generated as a result of Misrepair of DNA on multiple DNA breaks**

A SCC exhibits as deletion, translocation, or rearrangement of part of a chromosome. The altered part of a chromosome is often a DNA fragment that has been cut off from another chromosome by one or two DNA breaks. Chromosome translocation can be generated when a drop-off DNA fragment attaches incorrectly to another chromosome, which has also DNA breaks. Chromosome rearrangement can take place when a cut-off DNA fragment is reversed or recopied before being re-inserted back. Thus, generation of SCC is a consequence of DNA breaks. When there is only one DNA break in a cell, a PDM or deletion of part of a chromosome may occur. When there are multiple DNA breaks, PDM(s) and SCC(s) can be both generated. In fact, most cells that have multiple DNA breaks will die from failure of DNA. Survival of a cell with a PDM or a SCC is a rare case.

Since a PDM is generated as a result of Misrepair of DNA, generation of SCC should be also a result of Misrepair of DNA (Iliakis, 2015). Structural integrity of a DNA is essential for DNA function and for cell survival. Translocation or rearrangement of part of a DNA should be a result of re-linking of two DNA fragments. However, DNA re-linking can be achieved only by the DNA repair system of the cell, thus it is a result of DNA repair. DNA re-linking can be made by a direct end-rejoining of two DNA fragments, which may result in generation of a PDM. However, DNA re-linking can be also made via a "foreign" DNA fragment, which may result in generation of a SCC. The "foreign" DNA fragment can be: a translocated/inverted/recopied DNA fragment or a viral DNA fragment. Although part of DNA sequence is altered by such an incorrect re-linking of DNA, DNA integrity is maintained. Thus, a SCC is generated for cell survival. Insertion of a viral DNA into a host DNA is a type of SCC, thus it is also a result of Misrepair of DNA.

### 7.4 Numerical CC (NCC): generated as a consequence of dysfunction of cell division

NCCs such as hyperdiploid and hypodiploid are often seen in ALL cells. Generation of a NCC is a consequence of dysfunction of separation of homologous chromosomes during cell mitosis. This may occur when a dividing cell is injured. Most of the cells that are severely injured during cell division will die; however a few may survive with or without chromosome changes. A cell that is injured during cell mitosis can have different destinies in different situations. For example, if a cell is injured at phase S/G2 and DNA synthesis is interrupted, the cell will die. If a cell is injured at M-metaphase, and chromosome separation but not cell body division is affected, a daughter cell can be born with gain or loss of chromosomes. If a cell is injured at M-telophase, and chromosome separation is not affected but cell body division fails, the cell will have double sets of chromosomes (tetraploid). Since gain or loss of one or more chromosomes may affect multiple genes, most of the daughter cells that are born with abnormal number of chromosomes will die. However, rarely, certain forms of NCCs can be non-fatal for a LC. Such forms of NCCs do not cause cell death but rather drive cell transformation. The aneuploid in ALL is an example. Importantly, in the organs of hematopoiesis (in marrow) and activation of lymphocytes (in LNs/LTs), a large number of developing cells are constantly in processes of cell division. These cells have high risk of injuries during cell division by damaging factors, thus they have a risk of generation of NCCs.

A PDM/SCC is made for DNA repair for cell survival, thus generation of PDM/SCC is not really a mistake. Differently, a NCC is not generated for "repair"; thus survival of a cell with a NCC is a real mistake. Importantly, a NCC can affect multiple genes and may cause cell transformation rapidly. If such a mistake is not removed, the NCC may lead to death of the organism by causing cancer development. Some ALLs may develop as a consequence of such a mistake, because the cell transformation in these ALLs is made by aneupoid of chromosomes.

## 7.5 Pathogen-infection: as the main external source of DNA breaks for LCs in LNs/LTs

Generation of PDM/SCC is a consequence of DNA breaks. The DNAs in a cell are protected in nucleus. Each DNA is hidden in its supercoli structure, which appears as a chromosome at phase M and as heterochromatin at phase G0. Only the activated parts of a DNA are temperately in relax (extended) state and not in supercoli structure. Thus, in a cell, DNAs have lower risk of injury than other parts of the cell. However, DNAs are sensitive to radiation, some types of chemicals, and viral. For the LCs in a LN/LT, infections of viral and bacterial are the main sources of DNA damage (Figure 4). Bacterial can injure host DNAs indirectly by extoxin and endotoxin. Viral have higher impact on host DNAs than bacterial. Viral can enter, hide, and proliferate in host cells, thus they can attack host DNAs directly. The viral in large numbers in a host cell can cut host DNAs frequently at different points. Not protected in supercoli structure, the activated parts of a DNA are more sensitive to viral-attacking than other parts. All the LCs that have passed a LN/LT may have a risk of DNA breaks by viral attacking. Differently, the developing LCs in marrow cavity and in thymus, including lymphoblasts and pro-lymphocytes, may have lower risk of DNA breaks.

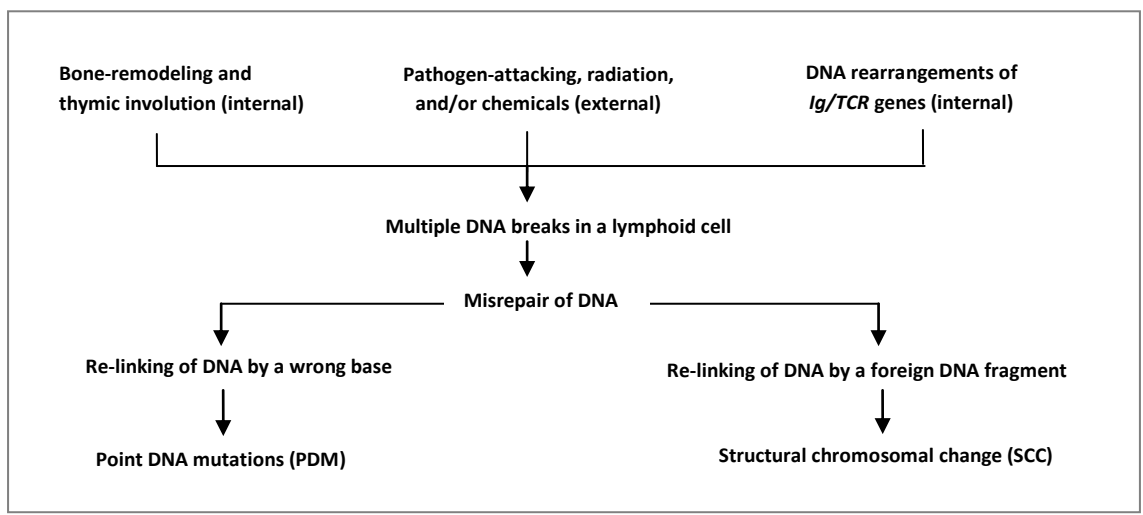

Figure 4. A PDM/SCC is generated as a consequence of DNA breaks in a cell

DNA breaks are the triggers for generation of PDM or SCC. Apart from pathogen-attacking and radiation/chemicals, bone-remodeling and thymic involution may be two internal sources of DNA breaks for LCs. DNA rearrangement of *Ig/TCR* genes may also introduce DNA breaks in a LC. A broken DNA must be re-linked for maintaining the structural integrity of DNA and for preventing cell death. However, if a broken DNA is re-linked by a wrong base, a PDM can be generated. If a broken DNA is re-linked by a foreign DNA fragment, a SCC can be generated. Thus, a PDM/SCC is generated as a consequence of DNA breaks in a cell.

## 7.6 DNA rearrangements of *Ig/TCR* genes may also introduce DNA breaks in a LC

DNA breaks are the triggers for generation of PDM or SCC. However, DNA breaks can be introduced in LCs not only by external damage but also possibly by DNA rearrangements of *Ig/TCR* genes. For producing immune cells that have large diversity on recognizing antigens, a lymphocyte undergoes three processes of "treatments" of *Ig/TCR* genes before becoming an immune cell. These three processes are: **A.** DNA rearrangement of V (D) J segments for the genes that code the variable regions of IgH/TCR, occurred in lymphoblasts; **B.** somatic hypermutation in the genes that code the variable regions of IgH, occurred in centroblasts; and **C.** class-switching recombination of DNA segments for the genes that code the Fc region of IgH, occurred in B-immunoblasts (Davila, 2001; Dadi, 2009). In processes **A** and **C**, certain DNA segments in loci of *IgH/TCR* genes needs to be cut off at first, and then the remaining parts of DNA need to be re-linked. Thus, in processes **A** and **C**, DNA breaks are introduced transiently (Figure 4).

In process **B**, hypermutation begins by modification of a base-C into a base-U in the locus of genes for the variable region of IgH. The "head" of base-U is then removed, leaving a non-paired point "facing" to the base-G in complementary strand. In subsequent DNA replication, this non-matching point will be paired by base N (A, T, C, or G). By this process, a base-C at this point in a mother cell is randomly replaced by A, T, C, or G in grand-children cells. However, in some cells, the locus of *IgH* gene have still non-paired base at this point. A non-paired base makes separation of two complementary strands of DNA at this point, which may affect the structural stability of the DNA. Thus, somatic hypermutation may make the part of DNA coding IgH have increased risk of DNA breaks.

Normally, the DNA breaks introduced in these processes can be closed quickly and correctly. However, if a cell is injured during one of these processes, incorrect re-linking of DNAs may take place, and a PDM or a SCC can be generated. For example, the t(8;14) in BL is possibly generated as a consequence of three DNA breaks in a centroblast (Cai, 2015). One DNA break (break A) occurs to the locus of *IgH* gene in chromosome 14, which may be related to somatic hypermutation of *Ig* gene. Two other breaks (break B and C) occur to chromosome 8, which may be related to viral attacking. If, by chance, DNA breaks B and C result in releasing of a DNA fragment containing *MYC* gene, MYC-translocation may occur. If the *MYC*-containing DNA fragment is unfortunately inserted into the locus of *Ig* gene by DNA break A, t(8;14) can be generated. Notably, a DNA translocation to the locus of *Ig/TCR* gene can have severe consequences. In a LC, *Ig/TCR* genes are constantly expressed. Thus, a gene that is translocated to the locus of an *Ig/TCR* gene can be constantly expressed or over-expressed by the promoter of *Ig/TCR* gene. Most LCs may be not able to survive from such a translocation.

For example, the IgH-MYC translocation in t(8;14) results in over-expression of MYC, which may drive the cell transformation of a centroblast in BL development. In T-LBL, the most frequent DNA changes are the translocations of transcription factor genes to the loci of *TCR* genes, such as t(7;9), t(1;14), and t(10;14). Such a translocation results in over-expression of a transcription factor and independent cell proliferation. A translocation to the locus of a *TCR* gene is quite possibly related to the DNA rearrangement of *TCR* genes during the differentiation of T-lymphoblast.

Taken together, there may be three groups of sources of DNA breaks for LCs: **A**. external radiation/chemicals and pathogen-attacking; **B**. internal bone-remodeling and thymic involution; and **C**. DNA rearrangements of *Ig/TCR* genes (internal) (Figure 4). Reparation of DNA breaks results in generations of PDMs and/or SCCs in LCs. Thus, repeated bone-remodeling, thymic involution, and pathogen-infections may all contribute to the enlargement of the reservoir of DNA changes in LCs with age of the body.

### 7.7 Repeated cell injuries and cell proliferations drive the accumulation of DNA changes in LCs

It is rare that "one" DNA change promotes cell transformation. Cell transformation is often a result of accumulation of many driver DNA changes. Most PDMs and some tiny SCCs are silent or mild to a cell, thus they can survive and accumulate in cells. A PDM/SCC is generated as a result of Misrepair of DNA on DNA breaks. However, a somatic cell has quite low opportunity to survive two times from severe DNA injuries. Thus, DNA changes cannot accumulate in "the same" cell, but only possibly in some of its offspring cells. Accumulation of Misrepairs of DNA needs to proceed in many generations of cells (Wang-Michelitsch, 2015).

Accumulation of DNA changes is a result of repetition of a circle of three processes: cell injuries and DNA injuries; survival of one of the injured cells by Misrepair of DNA and generation of PDM/SCC; and proliferation of this cell. Thus, accumulation of DNA changes is a slow process through many generations of cells. Repeated cell injuries are the triggers for generation of DNA changes, and repeated cell proliferation enables the accumulation of DNA changes in cells. For example, in a LN/LT, pathogen-attacking can trigger generation of DNA changes in a LC, and repeated proliferation of LCs promoted by pathogens enables the accumulation of DNA changes in memory cells and other LCs. In marrow cavity, repeated bone-remodeling may trigger generation of DNA changes in a HSC or a LC, and regeneration of HSCs and proliferation of developing LCs enable the accumulation of DNA changes in HSCs and LCs.

### 7.8 Long-term accumulation of DNA changes occurs mainly in HSCs and memory cells

DNA changes can accumulate in HSCs and LCs at any developing stage. However, a long-term accumulation of DNA changes can occur only in long-living stem cells. HSCs are stem cells regenerable for our whole lifetime. Memory cells are long-living immune cells that can be regenerated at each time of secondary immune response. Thus, among all DNA changes, only those generated or inherited in HSCs and memory cells can accumulate for a long time. The DNA changes in other LCs, if not inherited by memory cells, will disappear by death of the cells. For example, a DNA change generated in an effector cell will disappear when the cell dies.

Although accumulation of DNA changes takes place mainly in HSCs and memory cells, all of their offspring cells can inherit the DNA changes. A LC can have all the DNA changes of its precursor stem cells and other precursor LCs. For example, a naive lymphocyte can inherit the DNA changes generated in its precursor HSCs, lymphoblasts, and pro-lymphocytes. An effector cell can inherit the DNA changes generated its precursor HSCs, memory cells, and other precursor LCs. The accumulated DNA changes in HSCs and memory cells may contribute to cell transformation of any one of their offspring cells. Thus, the driver DNA changes for lymphoma development may be partially produced in the HSCs/LCs in marrow and partially in memory cells and LCs in LNs/LTs.

### 7.9  Accelerated accumulation of DNA changes in transformed LCs

Many forms of CCs (cytogenetic changes) and PDMs (molecular mutations) have been observed in leukemia cells and lymphoma cells. However, a big part of them might be only secondary DNA changes. Namely, they are the DNA changes that are generated after the occurrence of cell transformation of a normal cell into a tumor cell. Secondary DNA changes are not responsible for the occurrence of a cancer, but they may drive the further procession of a cancer and affect cancer prognosis.  Secondary DNA changes are often seen in late stage of a cancer. For example, gain or loss of an arm of chromosome, such as (+1q), (+7q), del (6q), del (13q), and del (17p), is often seen in later stage of MCL, chronic lymphocytic leukemia (CLL), and diffuse large B-cell lymphoma (DLBCL) (Royo, 2011; Grange, 2017). Complex karyotypes may be often secondary DNA changes in some lymphomas (Cohen, 2015). Generation of a secondary DNA change in a tumor cell is also a consequence of cell injuries and DNA injuries. However, the injuries of tumor cells may be often caused by the necrosis of some cancer cells due to lack of nutrition.

Importantly, accumulation of secondary DNA changes in neoplasm cells is self-accelerating and in-homogenous (Wang-Michelitsch, 2015). Three factors may contribute to this self-acceleration: **A.** frequent cell divisions of neoplasm cells; **B.** chromosomal instability in neoplasm cells caused by DNA changes; and **C.** increased tolerance of neoplasm cells to DNA changes because of poor cell differentiation. However, different neoplasm cells may have different secondary DNA changes, which lead to the heterogeneity of cancer cells. The cancer cells having different genetic backgrounds can be different to each other by morphology, behavior, and sensitivity to cancer therapy. This explains the treatment-resistance in late-stage cancers. Some secondary DNA changes can transform a well-differentiated neoplasm cell into a poor-differentiated one, which results in transformation of an indolent tumor into an aggressive cancer. In a tumor, if some tumor cells grow faster than others by some secondary DNA changes, they may become the so-called "tumor stem cells" (Martinez-Climent, 2010). A "tumor stem cell" may proliferate more quickly than other tumor cells; however, only a small group of cells in a tumor are produced by this "tumor stem cell". Therefore, a targeted treatment that takes "the tumor stem cells" as killing target cannot be very effective.

### VIII. Conclusions

We have discussed in this paper the potential sources of cell injuries of LCs and the mechanism of generation and accumulation of DNA changes in LCs. The DNA changes in LCs contributing to development of LL/lymphoma can be generated in marrow, thymus, and/or LNs/LTs. In LNs/LTs, pathogen-infections may be the main cause for cell injuries of LCs. In marrow cavity, repeated bone-remodeling during bone-growth and bone-repair may be a source of cell injuries of HSCs and developing LCs. In thymus, long-term thymic involution may produce damage to the developing T-cells. Generation of DNA changes is often a result of Misrepair of DNA on DNA breaks. DNA breaks can be introduced not only by external damage but also possibly by DNA rearrangements of *Ig/TCR* genes. Repeated cell injuries and repeated cell proliferation drive the accumulation of DNA changes in HSCs and LCs. Long-term accumulation of DNA changes occurs mainly in HSCs and memory cells.

To verify our hypothesis that bone-remodeling and thymic involution may be related to cell transformations of developing LCs, experimental researches can be undertaken to study respectively the associations of repeated bone injuries and repeated thymic injuries with the development of leukemia in animal models (such as rabbits).

It is sad when LL/lymphoma occurs in a child or an adolescent. Our analysis indicates that development of pediatric LL/lymphoma may be related to bone-growth and thymic involution. However, these two risk factors are unfortunately internal and unavoidable. In addition, immunodeficiency is a risk factor for both children and adults. Thus, to reduce the risk of occurrence of LL/lymphoma, we have two suggestions: avoiding violent sports and reducing the usage of anti-inflammation medicaments. Cryo-preserving the HSCs in umbilical cord at birth is advised for the whole population.